\documentclass[aps,pre,twocolumn,groupedaddress,floatfix,longbibliography,superscriptaddress]{revtex4-1}
\usepackage{amsmath}
\usepackage{amssymb}
\usepackage{amsfonts}
\usepackage{graphicx}
\usepackage{bbold}
\usepackage{bm}
\usepackage{bm}
\usepackage{cancel}
\usepackage{times,float}
\usepackage{graphicx}
\usepackage[usenames,dvipsnames,svgnames]{xcolor}
\usepackage{hyperref}
\hypersetup{colorlinks=true, linkcolor=NavyBlue, citecolor=PineGreen,urlcolor=cyan}
\usepackage{multirow}
\usepackage{braket}
\usepackage{float}
\usepackage{diagbox}
\usepackage{epsfig,subfig}
\usepackage{tabularx}
\usepackage{changepage}
\usepackage{soul}

\begin{document}

\title{Testing the equivalence principle with time-diffracted free-falling quantum particles}

\author{Juan A. Ca\~{n}as}
\affiliation{Universidad Ju\'{a}rez Aut\'{o}noma de Tabasco, Divisi\'{o}n Acad\'{e}mica de Ciencias B\'{a}sicas, 86690 Cunduac\'{a}n, Tabasco, M\'{e}xico}

\author{J. Bernal}
\affiliation{Universidad Ju\'{a}rez Aut\'{o}noma de Tabasco, Divisi\'{o}n Acad\'{e}mica de Ciencias B\'{a}sicas, 86690 Cunduac\'{a}n, Tabasco, M\'{e}xico}

\author{A. Mart\'in-Ruiz}
\email{alberto.martin@nucleares.unam.mx}
\affiliation{Instituto de Ciencias Nucleares, Universidad Nacional Aut\'{o}noma de M\'{e}xico, 04510 Ciudad de M\'{e}xico, M\'{e}xico}

\begin{abstract}
The equivalence principle of gravity is examined at the quantum level using the diffraction in time of matter waves in two ways. First, we consider a quasi-monochromatic beam of particles incident on a shutter which is removed at time $t = 0$ and fall due to the gravitational field. The probability density exhibits a set of mass-dependent oscillations which are genuinely quantum in nature, thereby reflecting quantum violations to the weak equivalence principle,  although the strong equivalence principle remains valid. We estimate the degree of violation in terms of the width of the diffraction-in-time effect. Second,  motivated by the recent advances in the manipulation of ultracold atoms and neutrons as well as the experimental observation of quantum states of ultracold neutrons in the gravitational field above a flat mirror, we study the diffraction in time of a suddenly released beam of particles initially prepared in gravitational quantum bound states. In this case, we quantify the degree of violation by comparing the time of flight from the mean position of the initial wave packet versus the time of flight as measured from the mirror.  We show that, in this case both the weak and strong versions of the equivalence principle are violated. We demonstrate that compatibility between equivalence principle and quantum mechanics is recovered in the macroscopic (large-mass) limit. Possible realizations with ultracold neutrons, cesium atoms and large molecules are discussed.
\end{abstract}

\maketitle

\section{Introduction} \label{Intro}

The equivalence principle is a cornerstone in the foundations of Einstein's general relativity, and it is indeed more fundamental than the symmetry of general covariance, in the sense that there are generally covariant spacetimes that do not obey the principle of equivalence  \cite{R_Wald_GR}. Essential to the search for a quantum theory  of gravity is a deep understanding of exactly where quantum mechanics and general relativity conflict. This motivates the search for tests of the equivalence principle of general relativity using quantum systems.

Different statements of the equivalence principle can be found in the literature. They correspond to propositions about  (i) the equality between inertial and gravitational masses, (ii) the universality of free fall, and (iii) the equivalence between homogeneous gravitational fields and uniform accelerated motion.

Several tests have been performed with pendula or torsion balances leading to extremely accurate confirmation of the equality of gravitational and inertial masses at the classical level \cite{C_Will}. It has also been proved with quantum-mechanical particles by using gravity-induced interference experiments \cite{PhysRevLett.34.1472, Peters1999}. Therefore, on the basis of these experimental demonstrations, in this paper we take for granted the validity of the statement (i).

The universality of free fall, often referred as the weak equivalence principle, asserts that all test bodies fall in a gravitational field with the same acceleration regardless of their mass or internal composition, provided they are small enough that one can neglect the effects of gravity gradients. This is exactly true in classical mechanics, as it is equivalent to the statement (i).  However,  in quantum systems, the universal character of gravity entails much more than just the equality between inertial and gravitational masses. Indeed, quantum objects do not satisfy the essence of the weak equivalence principle since their behavior in external gravitational fields (and even for free particles) is mass-dependent. This is clearly seen from the fact that while masses cancel out in the Newton's law for a particle in a homogeneous gravitational field (e.g. along the $z$-axis), $m \ddot{z} = m g$, they do not from the Schr\"{o}dinger equation,
\begin{align}
i \hbar \frac{\partial \psi}{\partial t} = - \frac{\hbar ^{2}}{2m} \frac{\partial ^{2} \psi}{\partial z ^{2}} + m gz \psi , 	\label{Schro-GravField}
\end{align}
thus implying that different inertial masses may produce different  spreading of wave packets. For a sharply peaked wave packet, by invoking Ehrenfest's theorem it is clear that the mean position $\braket{z}$ follows a geodesic, with however, mass-dependent quantum fluctuations around it (proportional to the ratio $\hbar / m$), thus signaling the nonuniversality of quantum free fall \cite{PhysRevD.55.455, Ali_2006}. The compatibility between the weak equivalence principle and quantum mechanics is an interesting issue that is yet to be completely settled. It has been extensively investigated with quantum particles that undergo free-fall in a homogeneous gravitational field. For example, in Ref. \cite{PhysRevD.55.455}, Viola and Onofrio studied free-falling Schr\"{o}dinger cat states and determine the average time of flight by means of the Ehrenfest's theorem, which unsurprisingly is mass-dependent. Also, violations to the weak equivalence principle has been investigated with Gaussian \cite{Ali_2006} and non-Gaussian \cite{Chowdhury_2011} wave packets evolving in the presence of the gravitational field, and quantified through the mean arrival time at an arbitrary detector location from a probability current approach \cite{PhysRevA.47.85, PhysRevA.51.2748, PhysRevA.59.1010, PhysRevA.58.840, LEAVENS199327, MUGA1995351}. On a different framework, in Refs. \cite{Davies_2004, Davies2_2004}, Davies used a model quantum clock \cite{PhysRev.109.571, doi:10.1119/1.12061} to compute the transit time of a free falling quantum particle in a background gravitational field. Recently, it has been proposed that violations to the weak equivalence principle can also be studied through the dephasing and phase shift of free-falling composite systems \cite{Anastopoulos_2018}. In these scenarios, a precise definition of the time of flight is required, and as discussed by Finkelstein in Ref. \cite{PhysRevA.59.3218}, there exists no unique or unambiguous definition that is universally applicable and  empirically well tested \cite{MUGA2000353}.

As a sequel to the works mentioned above, in this paper we study the issue of violations to the weak equivalence principle from a different perspective which allows us to avoid the phenomenological definitions of the time of flight, namely, the cut-the-wave procedure. This method consists in cutting up the wave function abruptly and evaluate the flux afterwards. By means of the peculiar time-dependence of the flux at a given position, i.e.,  diffraction in time, we quantify the degree of violation to the equivalence principle when the system evolves in the presence the gravitational field. This may be analyzed in terms of the Moshinsky shutter \cite{PhysRev.88.625}, who studied the time-evolution of a cut-off plane wave in free-space. Since its experimental verification \cite{PhysRevLett.77.4}, the diffraction in time phenomena has found many applications (see Ref. \cite{DELCAMPO20091} and references therein), and it has been even proposed as a probe for Planck scale physics \cite{PhysRevLett.101.221301, PhysRevD.90.125027}. 

The cancellation of masses in the Newton's law for a particle in a homogeneous gravitational field allow us to introduce a  coordinate system in which the accelerated motion with respect to an inertial reference frame is replaced by free motion in an accelerated frame,  thus implying the statement (iii), often referred as the strong equivalence principle.  In short,  performing the transformations $z^{\prime} = z - vt - \frac{1}{2} gt ^{2}$ and $t^{\prime}=t$,  the equation of motion $m \ddot{z} = m g$ for a particle in a homogeneous gravitational field reduces to the equation of motion for a free particle, i.e.,  $m \ddot{z} ^{\prime} = 0$.  In this way,  Einstein postulated that all the physical laws in a homogeneous gravitational field should be locally equivalent to the physics in a uniformly accelerated frame.  The question naturally arises as to whether or not the strong equivalence principle is valid for quantum systems.  The answer is in the affirmative, and it can be demonstrated by the fact that Schr\"{o}dinger's equation for a particle in a homogeneous gravitational field (\ref{Schro-GravField}) gets transformed, via the above coordinate transformation to an accelerated frame, to the free-particle Schr\"{o}dinger's equation:
\begin{align}
i \hbar \frac{\partial \psi ^{\prime}}{\partial t^{\prime}} = - \frac{\hbar ^{2}}{2m} \frac{\partial ^{2} \psi ^{\prime}}{\partial z ^{\prime \,  2}} , 	\label{Schro-Free}
\end{align}
with $\psi ^{\prime} = e ^{i \gamma (z,t) } \psi $. Therefore,  there is a formal correspondence between a uniform gravitational field and a uniform acceleration in the underlying quantum kinematics,  just as there is in classical kinematics.  However,  the relation between quantum states and its dynamical evolution is much more subtle than in the classical case.  In fact,  the equation of motion may transform correctly,  but the energy eigenstates do not i.e., while the energy eigenstates for a particle moving freely have the form $e ^{ikz-iEt / \hbar}$,  the stationary states of the Schr\"{o}dinger equation (\ref{Schro-GravField}) have the form $\mbox{Ai} (\frac{z - b}{a}) e ^{-iEt / \hbar}$, and they do not transform into each other under any coordinate transformation. In Ref. \cite{Longhi:18}, using entangled photons, violation of the strong equivalence principle was found, thus suggesting that we cannot take for granted its validity.

In this paper we examine the equivalence principle of gravity at the quantum level by using the diffraction in time of matter waves in two ways.  In Sec. \ref{DIT} we study a quasi-monochromatic beam of particles incident on a shutter that is suddenly removed at $t=0$ and fall due to the gravitational field.  The oscillatory behavior around the classical distribution is mass-dependent,  and we interpret it as a signature of the violation of the weak  equivalence principle.  The width of the diffraction in time serves a measure of the degree of violation.  We also show the validity of the strong equivalence principle in this case.  Next,  using the recent advances in the manipulation of ultracold atoms and neutrons,  as well as the experimental observation of quantum states of ultracold neutrons in the gravitational field above a flat mirror,  in Sec. \ref{GravityQuantumStates} we investigate the diffraction in time of a suddenly released beam of particles initially prepared in gravitational quantum bound states.  In this case,  since the quantum state is normalizable (as compared with the quasi-monochromatic beam),  we quantify the degree of violation in a different manner.  Indeed,  we compare the time of flight from the mean position of the initial distribution versus the classical time of flight.  We estimate the degree of violation for thermal and ultracold neutrons, as well as cesium atoms and large molecules.  In this case,  we find that the strong equivalence principle is violated.  In the last Sec. \ref{DiscussionConclusions} we discuss our results and present a brief summary.

\begin{figure}
\includegraphics[scale=0.5]{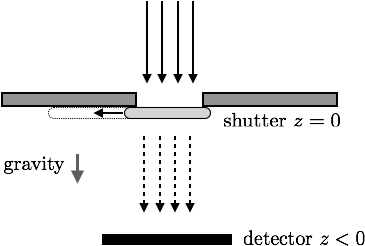}
\caption{The shutter problem.} \label{figure1}
\end{figure}

\section{Diffraction in time of free-falling particles}  \label{DIT}

Let us consider a monochromatic beam of independent particles (of mass $m$ and momentum $p > 0$) impinging on a totally absorbing shutter located at the origin $z=0$, such that at $t=0$, the shutter is opened and the beam is suddenly released in the presence of the gravitational field, as shown in Fig. \ref{figure1}.  We will refer to this setup as scenario A. This is the closest possible quantum version of the Galileo leaning tower experiment. The problem implies the following initial quantum state:
\begin{align}
\psi _{0} (z,t=0)=e ^{-i \frac{pz}{\hbar}} H (z) , \label{InitialStateDIT}
\end{align}
where $H(z)$ is the Heaviside step function. It is worth mentioning that the initial wave function (\ref{InitialStateDIT}) is idealized, since particles are not affected by gravity until the shutter is opened and they fall down. However this scenario can be achieved by applying a uniform electric field pointing upwards in the region $z>0$, compensating in this way the effects of gravity, making feasible the preparation of such initial state. In Section \ref{GravityQuantumStates} we shall consider a more realistic scenario where particles are trapped by the effect of gravity in bound states before they fall down.

\begin{figure}
\includegraphics[scale=0.45]{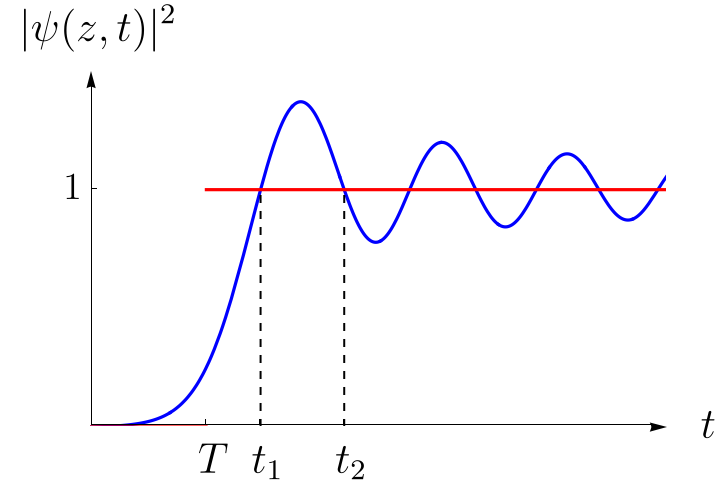}
\caption{Classical (red line) and quantum (blue line) density profiles as a function of time $t$ at a fixed distance $z<0$.} \label{DIT_Plot}
\end{figure}

The quantum state $\psi (z,t)$ at time $t>0$ is related with the initial quantum state (\ref{InitialStateDIT}) by
\begin{align}
\psi (z,t) = \int _{- \infty} ^{\infty} K (z,t ; z ^{\prime} , t ^{\prime}) \psi _{0} (z ^{\prime} , t ^{\prime}) d z ^{\prime} , \label{PropagatedStateDIT}
\end{align}
where $K \left( z ,t ; z ^{\prime} , t ^{\prime} \right)$ is the propagator, which solves the time-dependent Schr\"{o}dinger equation \cite{Feynman_Hibbs}.  For a uniform gravitational potential $V _{\mbox{\scriptsize Grav}} (z) = mgz$, the propagator is found to be
\begin{align}
\hspace{-0.2cm}  K \left( z ,t ; z ^{\prime} , t ^{\prime} \right) = \sqrt{\frac{m}{2 \pi i \hbar \left( t - t ^{\prime} \right)}} \exp \left\lbrace \frac{i}{\hbar} \left[ \frac{m \left( z - z ^{\prime} \right) ^{2}}{2  \left( t - t ^{\prime} \right)} \right. \right. \notag \\ \left. \left. - \frac{mg}{2} \left( t - t ^{\prime} \right) \left( z + z ^{\prime} \right) - \frac{mg ^{2}}{24} \left( t - t ^{\prime} \right) ^{3} \right] \right\rbrace . \label{FreeFallPropagator}
\end{align}
Leaving aside the technical details for evaluating the integral in Eq. (\ref{PropagatedStateDIT}), the time-evolved quantum state we obtain is
\begin{align}
\hspace{-0.2cm} \psi (x,t) &= \sqrt{\frac{1}{2}} e ^{i \phi (z,t)} \left\lbrace \left[ \frac{1}{2} + C (\xi)\right] + i \left[ \frac{1}{2} + S (\xi)\right] \right\rbrace , \label{PropagatedStateDIT2}
\end{align}
where $\phi (z,t)$ is a phase-factor irrelevant for the purposes of this paper, $C(\xi)$ and $S(\xi)$ are the Fresnel integrals \cite{Gradshteyn_Ryzhik}, and the Fresnel integral's argument is a function of position and time
\begin{align}
\xi = \sqrt{\frac{m}{\pi \hbar t}} \left( z + vt + \frac{gt ^{2}}{2} \right) , \label{Xi}
\end{align}
where $v=p/m$ is the particle's velocity. The quantum probability density $\vert \psi (z,t) \vert ^{2}$ admits a simple geometric interpretation in terms of the Cornu spiral (see for example Refs.  \cite{PhysRev.88.625, DELCAMPO20091, PhysRevD.90.125027}).  Figure~\ref{DIT_Plot} shows the classical and quantum probability densities as a function of time recorded by a detector located at $z<0$. While the classical distribution jumps suddenly from $0$ to the stationary value $1$ at $t=T$, where $T= - (v/g) + \sqrt{ (v/g) ^{2} + 2 \vert z \vert / g } >0$ is the (mass-independent) classical time of flight, the quantum distribution exhibits a diffraction effect in time: it increases monotonically from zero to $1/4$ for $t<T$ and it behaves as a damped oscillation around the classical value for $t>T$, tending to the exact classical value when $t \gg T$.

One of the main salient features of this setup is the validity of the strong equivalence principle.  As we can see,  the solution (\ref{PropagatedStateDIT2}) corresponds to that of the diffraction in time result in free-space \cite{PhysRev.88.625, DELCAMPO20091, PhysRevD.90.125027} via the coordinate transformation $z^{\prime} = z - vt - \frac{1}{2} gt ^{2}$ and $t^{\prime}=t$ to an accelerated frame.  On the other hand, as extensively discussed in Section \ref{Intro},  the possibility of violation of the weak equivalence principle in quantum mechanics has been largely discussed.  For example,  the mass dependence of both the probability density and the mean arrival time are taken as a convincing evidence of this. In quantum mechanics, the physical meaning of the mean arrival time $\tau$ has remained rather obscure. It is usually defined in terms of the quantum probability current density $J(z,t)$ by $\tau (z) = \frac{ \int _{0} ^{\infty} t \, J(z,t) \, dt  }{ \int _{0} ^{\infty} J(z,t) \, dt  }$, which is explicitly mass-dependent \cite{PhysRevA.47.85, PhysRevA.51.2748, PhysRevA.59.1010, PhysRevA.58.840, LEAVENS199327, MUGA1995351}. Another definition due to Peres is in terms of the expectation value of the clock-time operator $\hat{T}$, i.e., $\tau (z) = \bra{\psi (z,t)} \hat{T} \ket{\psi (z,t)}$, or equivalently, as a change in the phase of the wave-function \cite{PhysRev.109.571, doi:10.1119/1.12061}. 

To avoid referring 
to these phenomenological definitions of the mean arrival time, and taking advantage of the diffraction-in-time effect 
of our problem, here we quantify the degree of violation of the weak equivalence principle in terms of the width of the diffraction-in-time effect. 
It can be estimated from the difference $\delta t = t _{2} - t _{1}$ between the first two times at which the probability 
density takes the classical value, as shown in Fig.~\ref{DIT_Plot}.  Such times can be estimated from the Cornu spiral (see for 
example Refs. \cite{PhysRev.88.625, DELCAMPO20091, PhysRevD.90.125027}), with the result $\delta \xi \approx 0.85$.  For $p \vert z \vert \gg \hbar$ we obtain, from Eq. (\ref{Xi}),
\begin{align}
\delta t \simeq \delta \xi  \, \sqrt{ \frac{\pi v T}{k  \left( 2 \vert z \vert  - vT \right) ^{2}} } \; T .  \label{DeltaT}
\end{align}
Clearly, this quantity tends to zero in the limit of large-mass. This implies 
that diffraction-in-time effects vanish 
for macroscopic objects. As an example we estimate $\delta t$ for different probes. We first consider a neutron beam at a 
thermal energy of 
$0.0253$eV ($v \simeq 2200$m/sec) \cite{EMRICH201655} and the detector placed at $\vert z \vert =1$m. The diffraction width (\ref{DeltaT}) becomes
\begin{align}
\delta t = 0.37 \times 10 ^{-8} \mbox{sec} , \qquad \mbox{thermal neutrons} ,
\end{align}
which is very small. If instead we consider ultracold neutrons (UCNs), for which $v \simeq 2$cm/sec \cite{Nesvizhevsky2002}, the diffraction width results
\begin{align}
\delta t = 6 \times 10 ^{-5} \mbox{sec} , \qquad \mbox{ultracold neutrons} ,
\end{align}
which is four orders of magnitude larger. A similar order of magnitude is obtained by using cesium atoms \cite{PhysRevLett.71.3083}, for which $m \simeq 2.2 \times 10 ^{-25}$kg. The result is
\begin{align}
\delta t = 0.5 \times 10 ^{-5} \mbox{sec} , \qquad \mbox{Cesium atoms} .
\end{align}
Therefore, these results make UCNs 
and cesium atoms potential candidates to test violations to the weak equivalence principle, as well as the validity of the strong equivalence principle, through diffraction in time experiments.
 
Large molecules, such as C$_{60}$, C$_{176}$ and large organic molecules, have 
been proposed to be promising candidates for indirect probes of quantum gravity in a laboratory setting. Time diffraction effects are expected to be considerably larger than those of candidate theories for quantum gravity, so we expect them to be a promising alternative to test 
the equivalence principle according to our predictions. For example, C$_{60}$ \cite{Arndt1999} and C$_{176}$ \cite{Goel2004} buckyball molecules have masses $1.19668 \times 10 ^{-24}$kg and $3.50706 \times 10 ^{-24}$kg, respectively, and taking $v \simeq 2$cm/sec, we find
\begin{align}
\delta t = 0.4 \times 10 ^{-6} \mbox{sec} , \qquad \mbox{C$_{60}$ molecule} \\  \delta t = 0.18 \times 10 ^{-6} \mbox{sec} , \qquad \mbox{C$_{176}$ molecule} .
\end{align}

We close this section by commenting about the formal emergence of the classical limit in this system. As discussed before, as the mass increases, 
the time width (\ref{DeltaT}) decreases,  thus indicating the convergence to the classical time of flight.  Therefore,  we expect the quantum probability distribution to approach  its 
classical counterpart in the same fashion. It is widely accepted that classical and quantum probability density functions approach each other for periodic systems in a locally averaged sense when the principal quantum number is large (i.e., 
in the high-energy limit) \cite{doi:10.1119/1.17807,  doi:10.1119/1.2173280,  Rowe_1987}. This has been successfully confirmed for the simplest spatially confined quantum systems: the infinite square well potential, the harmonic oscillator, the Kepler problem \cite{ClassLim1, ClassLim2, ClassLim3},  and more recently the quantum bouncer \cite{Nuevo}. However, the application of this prescription to systems with continuous spectra is rather unclear. In the present case,  the oscillatory behavior of the quantum distribution due to time diffraction, together with the energy level controlled by the mass, allows  us to extend 
the idea of local averages to the time domain as follows: the local average in the time domain of the quantum probability density follows the classical distribution for large masses. In short:
\begin{align}
\rho _{\mbox{\scriptsize C}} (z,t) = \lim _{m \gg 1} \frac{1}{2 \epsilon _{m}} \int _{t - \epsilon _{m}} ^{t + \epsilon _{m}} \vert \psi (z,t ^{\prime}) \vert ^{2} dt ^{\prime} , \label{LocalAverage}
\end{align}
where the interval $\epsilon _{m}$ decreases as the mass $m$ increases.  In the large-mass limit, we use the asymptotic form of the Fresnel integrals \cite{Gradshteyn_Ryzhik} to evaluate this expression,  thus confirming the emergence of the classical distribution.

\section{Suddenly released gravitational quantum states}  \label{GravityQuantumStates}

The observation of gravitational quantum states of ultracold neutrons \cite{Nesvizhevsky2002} has opened a new arena in which new fundamental short-range interactions \cite{BAELER2009149} and physics beyond the Standard Model \cite{PhysRevD.97.095039, PhysRevD.99.075032} can be tested. Inspired by the  experiments performed with UCNs at the Institut Laue-Langevin, in this section we suggest that a time-diffracted beam of UCNs, initially prepared in gravitational quantum states, serve also as a probe for the validity of both the weak and the strong equivalence principles in the quantum regime. In order to gain a clear insight of the idea, we briefly recall how the GRANIT experiment works.

\begin{figure}
\includegraphics[scale=0.5]{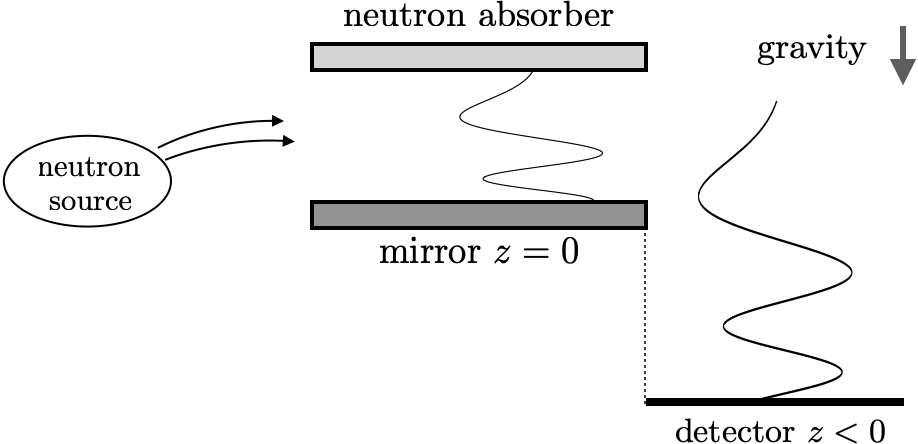}
\caption{Schematic of the setup.  Arrows correspond to neutron classical trajectories between the source and the entrance to the slit between the mirror (at $z=0$) and a neutron absorber. The oscillatory gray curve illustrates the 
squared moduli of the neutron wave function above the mirror, whose quantum state can be selected by varying the height of the neutron absorber.  In this way, neutrons are prepared in low-energy quantum states 
and then fall 
to the detector at $z<0$.} \label{figure2}
\end{figure}

In the experiment sketched in Fig. \ref{figure2}, an intense horizontal beam of UCNs is allowed to fall onto a horizontal mirror. By using a neutron absorber right above the mirror and counting the number of particles that move up to the absorber and down to the mirror, they found that UCNs do not move continuously but jump from one height to another, as quantum theory predicts. In this situation, the vertical motion is quantized, while the horizontal one is driven by classical laws. Here, we suggest that if the mirror is suddenly removed, or equivalently when UCNs reach the end of the chamber and freely-fall in the presence of the gravitational field,  as depicted in Fig. \ref{figure2}, the time-diffracted neutrons can be used as a probe to test the equivalence principle in an analogous way as the system considered in the previous section. Of course, the initial quantum state will be prepared inside the chamber, and hence, one can further explore the validity of the various forms of the equivalence principle in the quantum regime (for neutrons in low-energy states) as well as the transition to the classical regime (for neutrons in high-energy states). In principle, a similar experiment can be carried out using ultracold atoms.  We will refer to this setup as scenario B.

The neutron wave function $\psi (z)$ in the Earth's 
gravitational field above a mirror is governed by the Schr\"{o}dinger equation
\begin{align}
\left( - \frac{\hbar ^{2}}{2m} \frac{d ^{2} }{dz ^{2}} +mgz \right) \psi (z) = E \psi (z) ,  \label{Schro-GravUn}
\end{align}
subjected to the following boundary conditions: $\psi (z)$ must vanish asymptotically as $z \to \infty$,  and $\psi (z = 0) = 0$ because of the presence of a mirror at $z=0$.  All in all,  the normalized solution can be written in terms of the Airy function \cite{doi:10.1142/p345}
\begin{align}
\psi _{n} (z) = \frac{1}{\sqrt{l _{g}}} \frac{\mbox{Ai} (a _{n} + z/l _{g} )}{\mbox{Ai} ^{\prime} (a _{n})} H (z) , \label{UCNsWaveFunc}
\end{align}
where $a _{n}$ is the $n$-th zero of the Airy function $\mbox{Ai}$ and $l _{g} = \sqrt[3]{\hbar ^{2} / (2m ^{2}g ) }$ is the gravitational length. The boundary condition at $z=0$ defines the quantum state energies \cite{NESVIZHEVSKY2000754}
\begin{align}
E _{n} = - mgl_{g} a _{n} . \label{UnpEnergy}
\end{align}
Within the classical description, a neutron with energy $E _{n}$ can rise in the gravitational field up to the height $h _{n} = E _{n} / mg = - a _{n} l _{g}$. These idealized conditions are precisely the ones realized in the GRANIT experiment with ultracold neutrons in the Earth's gravity field \cite{Nesvizhevsky2002}.

Once the quantum states of the UCNs are prepared, they are allowed to freely fall due to gravity. Note that unlike the idealized initial quantum state considered in Section \ref{DIT}, the one considered here is perfectly feasible from the experimental side. As before, the time evolution of the state is governed by Eq. (\ref{PropagatedStateDIT}), with the propagator given by Eq. (\ref{FreeFallPropagator}) and the initial quantum state of Eq. (\ref{UCNsWaveFunc}), i.e. $\psi _{0} (z,t=0) = \psi _{n} (z)$. After simple manipulations, the time-evolved state can be expressed in the integral form
\begin{align}
\psi _{n} (z,t) &= \sqrt{ \frac{m  l _{g}}{2 \pi \hbar t}} \int _{a_{n}} ^{\infty} d \chi \; \frac{\mbox{Ai} ( \chi )}{\mbox{Ai} ^{\prime} (a _{n})} \notag \\ & \hspace{0.2cm} \times  \exp \left\lbrace i  \frac{m}{2 \hbar t}   \left( z - h _{n} + \frac{1}{2} g t ^{2}  - l _{g} \chi  \right) ^{2}   \right\rbrace  , \label{Time-Evolved-UCNsWaveFunc}
\end{align}
which cannot be expressed in an analytical closed-form. However, we appeal for numerical calculations in order to visualize the probability density and draw some conclusions.

\begin{figure}
\subfloat[\label{SmallMGround}]{\includegraphics[width = 1.7in]{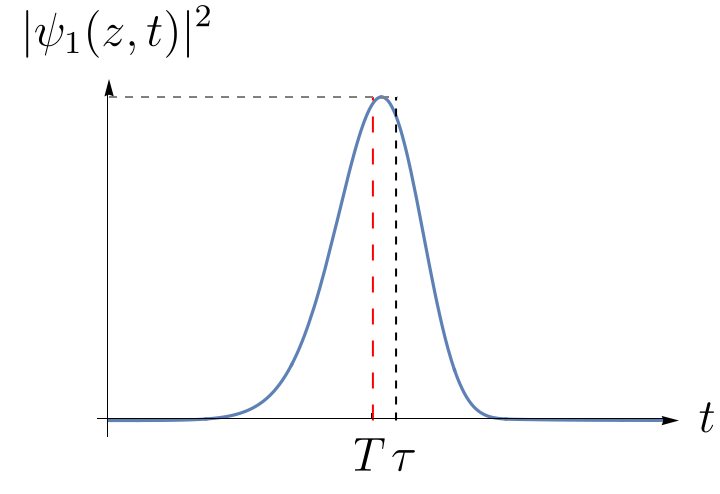}}
\subfloat[\label{LargeMGround}]{\includegraphics[width = 1.7in]{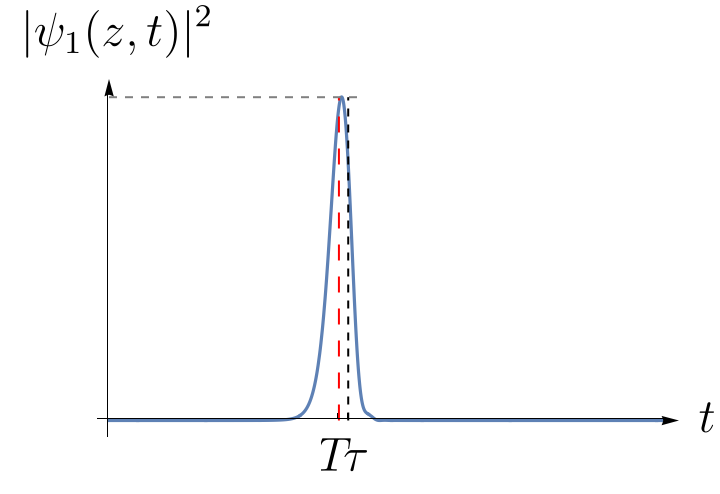}} \\
\subfloat[\label{SmallMExcited}]{\includegraphics[width = 1.7in]{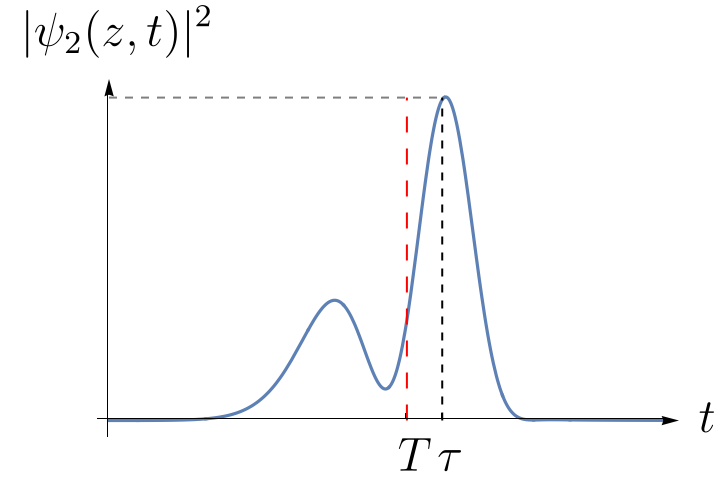}}
\subfloat[\label{LargeMExcited}]{\includegraphics[width = 1.7in]{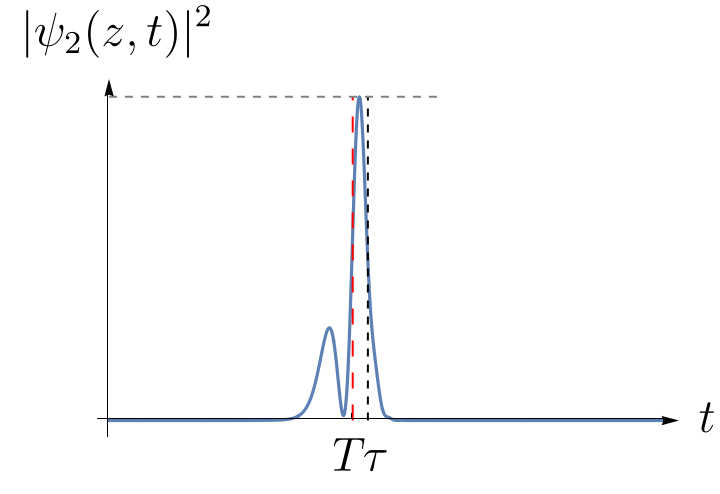}}
\caption{Plots of the quantum probability densities for ground state (upper panel) and the first excited state (lower panel), for small-mass (at left) and large-mass (at right).} \label{DensityProfiles}
\end{figure}

In Fig. \ref{DensityProfiles} we show the probability density $\vert \psi _{n} (z,t) \vert ^{2}$ as a function of time $t$ at a fixed distance $z<0$ for different initially prepared quantum states (\ref{UCNsWaveFunc}) and different masses. In the upper panel, we show the ground state ($n=1$) for a light mass (at left), a neutron for instance, and for a heavier object (at right) such as large molecules. The lower panel shows the first excited state ($n=2$), and as before, at left (right) we take a  light (heavy) mass.

There are some important differences with respect to the case presented in the previous section that deserves to be discussed in detail. The most salient feature is perhaps the normalizability of the quantum states. As we can see in the Fig. \ref{DIT_Plot}, the probability of finding time-diffracted particles after the shutter has been removed, i.e., the area under the curve on the interval $[0, \infty )$, is infinite. This is due to the fact that the initial state (\ref{InitialStateDIT}) is not really monochromatic since the spatial truncation, besides it is clearly not normalized. This is what makes the probability density to oscillate about the classical value for any time $t$ considerably larger than the time of flight $T$. On the other hand, as evinced in Fig. \ref{DensityProfiles}, the probability of finding a particle initially prepared in a gravitational quantum state (\ref{UCNsWaveFunc}) is finite. This is due to probability conservation, since the initial quantum state (\ref{UCNsWaveFunc}) is properly normalized. We now turn to the interpretation of our results.

It is well-known that the gravitational quantum state (\ref{UCNsWaveFunc}) exhibits oscillations from the ground level at $z =0$ up to infinity, and the number of nodes is determined by the quantum number $n$. In the GRANIT experiment, a series of quantized heights $h _{n} = - a _{n} l _{g}$ are measured \cite{Nesvizhevsky2002}, which correspond to the classical turning points, i.e., they measure the population of the $n$th quantum state of the UCNs. In Fig. \ref{DensityProfiles} we observe that the time-evolved probability density exhibits the same nodes as the initial quantum state, independently of the mass. From Eq. (\ref{Time-Evolved-UCNsWaveFunc}) we read $z(t)= h _{n} - \frac{1}{2} gt ^{2}$ as a possible equation of motion in the large-mass limit, wherefrom we determine the time of flight $\tau = \sqrt{ \frac{2}{g} (\vert z \vert + h _{n}) }$. This corresponds to the classical free fall time from the classical turning point $h _{n}$. 

In the ground state, which is expected to exhibit non classical behavior, it is clear that the center of the quantum distribution is not peaked at the time $\tau$ (black-dotted vertical line) defined above, but at a lesser time $T$ (red-dashed vertical line), as shown in Figs. \ref{SmallMGround} and \ref{LargeMGround}. This occurs because from a classical point of view, we have to consider that the particle freely-falls not from the turning point $h _{n}$, but from the mean position of the initial distribution, which is $\braket{z} = \frac{2}{3} h _{n}$. Following this idea we introduce the time scale 
\begin{align}
T = \sqrt{ \frac{2}{g} \left( \vert z \vert + \frac{2}{3} h _{n} \right) } < \tau , 
\end{align}
which is in clear agreement with the center of the time-evolved quantum distributions shown in Figs. \ref{SmallMGround} and \ref{LargeMGround}. Due to the mass-dependence of the gravitational length $l _{g} \propto m ^{-2/3}$, it is clear that $T \sim \tau$ in the large-mass limit, as we can confirm in the plots since the vertical dotted and dashed lines approach each other as the mass increases. Even more, in the limit $m \to \infty$, these time scales converge to the Newtonian time of flight $t _{\mbox{\scriptsize class}} = \sqrt{ 2 \vert z \vert / g }$, since the initial quantum distribution becomes strongly peaked at $h _{n} \to 0$, i.e., $ \lim _{m \to \infty} \vert \psi _{n} (z) \vert ^{2} \propto \delta (z)$ and $\braket{p} = 0$ (vanishing initial velocity). Interestingly, this case corresponds to a beam of particles moving horizontally on the mirror and then reflected into vertical motion. This implies that the time-evolved quantum state is given by the propagator (\ref{FreeFallPropagator}) with $z^{\prime}=0$, and hence, no transient effect takes place. Our result supports that the validity of the equivalence principle emerges in the large-mass limit (or equivalently in the large-energy limit, recall that $E _{n}=mgh _{n}$), which is commonly taken for granted. The same qualitative behavior is displayed by excited states when the mass is increased, as we can see in Figs. \ref{SmallMExcited} and \ref{LargeMExcited}.

In the previous section we introduced the diffraction width (\ref{DeltaT}) to quantify the degree of violation of the weak equivalence principle. In the present case, the quantum distribution does not increase up to the classical value due to its finite width and spreading, so we cannot use the same quantity to measure the degree of violation of the weak equivalence principle. However, the quantum distribution is strongly peaked at a certain value of time $T$ at a fixed distance $z<0$, which is different from the Newtonian time of flight $t _{\mbox{\scriptsize class}}$. Therefore, we can use this time delay to quantify the departures from the classical behavior: for a detector at a fixed position $z<0$, the transit time of particles deviates from the classical result by
\begin{align}
\delta t = \frac{T - t _{\mbox{\scriptsize class}}}{t _{\mbox{\scriptsize class}}} = \frac{h _{n}}{3 \vert z \vert} , \label{TimeDelay}
\end{align}
which tends to zero when the mass goes to infinity. This ratio can be estimated in a simple fashion for different quantum systems. They are summarized in Table \ref{table}. As expected, for very light particles, the transit times (\ref{TimeDelay}) deviate strongly from the Newtonian value, but for heavier objects (such as a cesium atom or large molecules as C$_{60}$ and C$_{176}$) they approach, reassuringly, to the expected mass-independent classical result. 

\begin{table}[h!]
\centering
\begin{tabular}{|c|c|c|c|c|} 
 \hline
 $n$ & Neutrons & cesium & C$_{60}$ & C$_{176}$ \\ [0.5ex] 
 \hline
 1 & $4.6 \times 10 ^{-6}$ & $4.77 \times 10 ^{-7}$ & $5.72 \times 10 ^{-8}$ & $2.06 \times 10 ^{-8}$ \\
 2 & $8 \times 10 ^{-6}$ & $3.1 \times 10 ^{-7}$ & $1 \times 10 ^{-7}$ & $3.61 \times 10 ^{-8}$ \\ [1ex] 
 \hline
\end{tabular}
\caption{Time delay in the free fall of quantum systems initially prepared in gravitational quantum states.}
\label{table}
\end{table}

\begin{figure}
\subfloat[\label{LargeMGroundApp}]{\includegraphics[width = 1.7in]{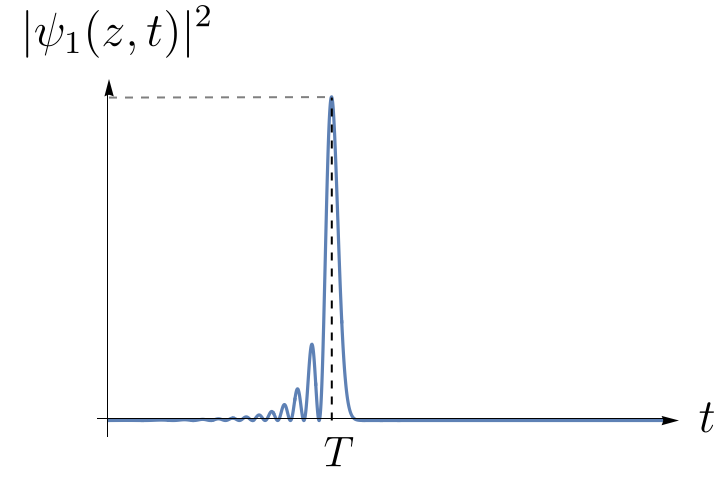}}
\subfloat[\label{LargeMExcitedApp}]{\includegraphics[width = 1.7in]{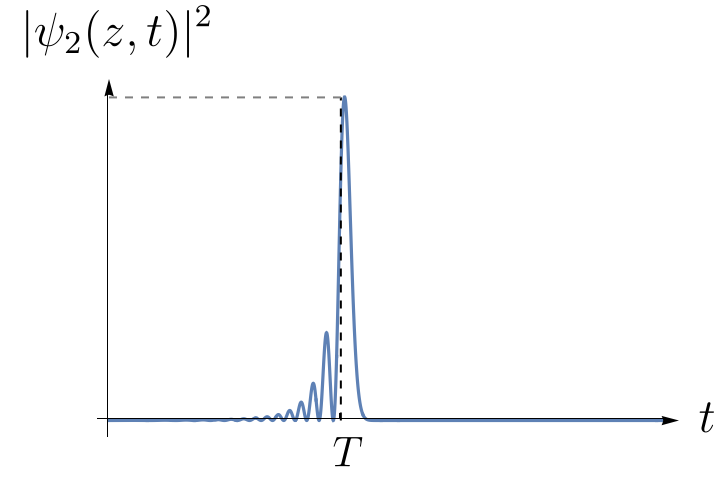}}
\caption{Plots of the quantum probability densities, approximated by the steepest descend method in the large-mass limit, for the ground state (at left) and the first excited state (at right).} \label{DensityProfilesApp}
\end{figure}

From the numerical analysis of the time of flight, it is clear that the classical behavior,  together with the validity of the weak equivalence principle,  emerges in the large-mass (high-energy) limit,  in agreement with the local averaging procedure (\ref{LocalAverage}) discussed in the previous section.  From the integral expression for the time-evolved quantum state (\ref{Time-Evolved-UCNsWaveFunc}), we observe by inspection that the large-mass limit makes the exponential a rapidly oscillatory function, while the Airy function in front smoothly oscillates since it is mass-independent. Since both functions are analytic in the complex plane, we can obtain an approximation of the integral in the large-mass limit by means of the steepest descend method. We observe that the exponential has a unique stationary point (i.e. where the integrand is maximum) at
\begin{align}
\chi _{0} (z,t) = \frac{1}{l _{g}} \left( z - h _{n} + \frac{1}{2} g t ^{2}  \right) . \label{Chi0}
\end{align}
We now expand the Airy function in Eq. (\ref{Time-Evolved-UCNsWaveFunc}) to the zeroth order (other terms are neglected) and the function in the exponential up to second order. Thus, the quantum state may be written as
\begin{align}
\psi _{n} (z,t) &= \sqrt{ \frac{m  l _{g}}{2 \pi \hbar t}} \frac{\mbox{Ai} ( \chi _{0} )}{\mbox{Ai} ^{\prime} (a _{n})} \int _{a_{n}} ^{\infty} d \chi  \exp \left[ i  \frac{m l _{g} ^{2}}{2 \hbar t}   \left(\chi _{0} - \chi  \right) ^{2}   \right]  . \label{Time-Evolved-UCNsWaveFuncApp}
\end{align}
This integral is quite simple. It can be expressed in terms of the Fresnel integrals as:
\begin{align}
\psi _{n} (z,t) &= \sqrt{ \frac{1}{2 l _{g}}} \frac{\mbox{Ai} ( \chi _{0} )}{\mbox{Ai} ^{\prime} (a _{n})} \left\lbrace \left[ \frac{1}{2} + C (\xi)\right] + i \left[ \frac{1}{2} + S (\xi)\right] \right\rbrace  , \label{Time-Evolved-UCNsWaveFuncApp2}
\end{align}
where $\chi _{0}$ is given by Eq. (\ref{Chi0}) and
\begin{align}
\xi (z ,t ) = \sqrt{ \frac{m}{\pi \hbar t} } \left( z + \frac{1}{2} g t ^{2} \right) .
\end{align}
In Fig. \ref{DensityProfilesApp} we plot the quantum probability density $\vert \psi _{n} (z,t) \vert ^{2} $, approximated by the steepest descend method in the large-mass limit (\ref{Time-Evolved-UCNsWaveFuncApp2}), for the ground and first-excited states. In both cases we observe a sharp peak which coincides very well with the time of flight $T$ defined above, and a series of secondary small peaks. 

We finally discuss both,  the emergence of the classical behavior as well as the validity of the strong equivalence principle.  The former is a simple task  in this case.  Using the limiting form of the Airy function $\delta (x) = \lim _{\epsilon \to 0 ^{+}} \frac{1}{ \epsilon } \mbox{Ai} (x/ \epsilon )$,  in the large-mass limit,  the wave function (\ref{Time-Evolved-UCNsWaveFuncApp2}) becomes
\begin{align}
\psi _{n} (z,t) \approx \delta  \left( z + \frac{1}{2} g t ^{2} \right)  ,  \label{ClassLim}
\end{align}
which is mass-independent and it is nonvanishing only along the exact classical trajectory. Of course, quantum corrections (proportional to the ratio $\hbar / m$) may arise to account for the subdominant quantum behavior of the freely falling 
particle in the classical limit.  The result of Eq.  (\ref{ClassLim}) can be understood in a simple fashion. In the large-mass limit, the gravitational length tends to zero, and hence,  for low-energy quantum states the mean position of the initial wave packet is at the ground level $z = 0$, and the mean velocity is zero because of the parity of the eigenstates. Therefore, in the classical limit, the particle freely-falls from the mirror and the corresponding equation of motion is $z = - gt^{2}/2$, as suggested by the wave function (\ref{ClassLim}).

There is just one question left to be answered: Is 
the strong equivalence principle valid or not for this quantum system? Our answer is in the negative, as we shall discuss.  Changing coordinates to an accelerated frame, i.e. $z^{\prime} = z - \frac{1}{2} gt ^{2}$ (with vanishing initial velocity) and $t^{\prime}=t$ , the Schr\"{o}dinger equation (\ref{Schro-GravUn}) transforms into the free-particle equation $\psi ^{\prime \prime} (z) = 0$, with however the time-dependent boundary condition $\psi ( - \frac{1}{2} gt ^{2})=0$ (i.e., 
the ground level freely-falls at the same rate that the particle does).  Nevertheless, the solution to this problem is quite different from the one presented in the exact time-evolved 
state (\ref{Time-Evolved-UCNsWaveFunc}), thus indicating that the strong equivalence principle is explicitly violated.  However, there is a limit in which this principle works again: the large-mass limit.  Using the steepest descend method, the exact time-evolved state (\ref{Time-Evolved-UCNsWaveFunc}) can be approximated by Eq. (\ref{Time-Evolved-UCNsWaveFuncApp2}), which we understand as composed by the product of two parts (leaving aside possible phase factors).  The first one corresponds to the initial quantum state (\ref{UCNsWaveFunc}) as seen from an accelerated frame, i.e.  
\begin{align}
\mbox{Ai} (a _{n} + z/l _{g} )  \quad \to \quad \mbox{Ai} \left( a _{n} + \frac{z + gt^{2}/2 }{l _{g}} \right) .
\end{align}
So, this part indicates that the initial quantum state evolves in time without distortion. However, the second part in Eq. (\ref{Time-Evolved-UCNsWaveFuncApp2}) has the information concerning the diffraction-in-time effect 
due to gravity, thus violating explicitly the statement (iii) for the strong equivalence principle.  This result confirms the finding in Ref.  \cite{Longhi:18} regarding the violation in a quantum simulation. However, in the large-mass limit, the Fresnel integrals approach $1/2$, thus converting this part into a phase factor. Therefore,
\begin{align}
\psi _{n} (z,t) \quad \to \quad \psi _{n} (z + gt^{2}/2 ) e ^{i \pi /4} ,
\end{align}
thus recovering the validity of the strong equivalence principle in this system.  This suggests that both the weak and strong versions of the equivalence principle are profoundly related in the quantum realm, just as they are in classical physics. The former is always violated due to the finite extension of wave packets,  thus activating the violation of the latter (even when the Schr\"{o}dinger equation may transform correctly), as our results show. Therefore, in the large mass-limit, both statements work again. As evinced by Eq. (\ref{TimeDelay}), large masses imply $\delta t \to 0$, thus recovering the classical time of flight.  Besides, the same limit leads to the validity of the strong version because the time-evolved state is obtained by transforming to an accelerated frame, in which the time-diffraction 
effects result in an mere irrelevant phase factor.

\section{Discussion and Conclusions}\label{DiscussionConclusions}

The universality of the ratio between the gravitational and inertial masses was established with the Galileo's famous gedanken experiment of dropping bodies of different mass from a great height. The general theory of relativity, and its plausible variants, are founded in the equivalence principle. One of its many faces is precisely the equality between the gravitational and inertial masses, and it has been confirmed experimentally both from the classical and quantum sides. A different statement of the equivalence principle refers to the universality of free-fall (also referred as the weak equivalence principle). In classical mechanics, it is perfectly equivalent to the equality between the gravitational and inertial masses, since they cancel out in the Newton's equation of motion, thus confirming the Galileo's conclusions. However, if Galileo's experiment is performed with quantum probes, the universality of free fall ceases to be valid since the mass does not cancel out from the Schr\"{o}dinger equations both, for a free particle and for a particle in a uniform gravitational field. The compatibility between the weak equivalence principle and quantum mechanics has been investigated with wave packets and quantified by means of the time of flight to a detector arbitrarily located. Unsurprisingly, the time of flight is mass-dependent, and converges to the classical result in the large-mass limit. In quantum mechanics, the definition and the physical meaning of the time of flight is rather obscure. Some works use a probability current approach to define a mean time, and others use a model quantum clock to define a transit time in terms of the variation of the phase of the wave function between two given positions.  

In this paper we have considered physical phenomena in which time naturally arises: the diffraction-in-time effect  It consists in a beam of particles suddenly released in the presence of the gravitational field, as shown in Fig. \ref{figure1}.  So, measuring the flux afterwards,  we can determine the transit time at a fixed detector.  We refer to this setup as scenario A. It is worth mentioning that diffraction in time (i.e.  the quantum distribution as a function of time) has been experimentally detected, so the configuration we propose is feasible to be realized in lab, and it can be used to test both the weak and strong versions of the equivalence principle. However as pointed out in Section \ref{DIT},  the initial wave function we consider is rather idealized, since it assumes that gravity is absent above the shutter.  This scenario can be achieved however by imposing a homogeneous electric field above the shutter to offset the effect of gravity,  and hence the plane-wave initial state can be prepared. In scenario A we find that the weak equivalence principle is violated (since the transit time depends on the mass) and the strong equivalence principle is valid (the time-evolved state is obtained by a coordinate transformation of the free case to an accelerated reference frame).  Using the width of the diffraction-in-time effect, we quantify the degree of violation to the weak  equivalence principle, and give some numerical estimates for thermal neutrons, ultracold neutrons, cesium atoms and large molecules, such as C$_{60}$ and C$_{176}$.  We find that ultracold neutrons and cesium atoms are the best probes to test our predictions,  since the width of the diffraction effect in time is of the order $10^{-5}$s, which is within the current experimental precision. 

Motivated by the recent high-sensitivity GRANIT experiments with ultracold neutrons,  in this paper we have also considered  the physics of ultracold neutrons as a test bed for studying violations to the equivalence principle. This configuration, sketched in Fig. \ref{figure2} and termed scenario B, is quite different from the previous one,  since the initial quantum state is normalizable, and hence the time-evolved quantum state does not exhibit the same time profile in the quantum distribution.  As such, we adopt a different mechanism to quantify the mass-dependence of the transit time.  We showed that the quantum time of flight, in the large-mass (high-energy) limit,  corresponds to the classical free fall time when the particle is suddenly released from the mean position, as computed with the initial quantum state.  As expected,  we find that in the large-mass limit,  the quantum transit-time converges to the classical time of flight.  We showed that this system violates both,  the weak and strong equivalence principle.  The former is expected due to the finite extent of the quantum distribution, however, the latter is subtle since we know that the Schr\"{o}dinger equation for a particle in a gravitational field correctly transforms into the Schr\"{o}dinger equation for a free-particle. Nevertheless,  this system shows that it is not the case with the energy eigenstates.

Finally, our conclusions (summarized in Table \ref{table2}) do not attempt to completely settle  the problem of the compatibility between the equivalence principle and quantum mechanics. We have considered two particular configurations in which time plays a prominent role, that serve as a quantum probe for both the universality of free fall and for the equivalence between homogeneous gravitational fields and uniform accelerated motion.  Both systems violate the weak equivalence principle,  as expected.  However,  when using a beam of quasi-monochromatic particles,  the strong equivalence principle holds,  while the case for quantum states bounded by the gravitational field,  it is violated.  This confirms that although Schr\"{o}dinger equation may transform correctly, the energy eigenstates do not.

\begin{table}
\centering
\begin{tabular}{|c|c|c|c|c|} 
 \hline
   & \quad Scenario A \quad & \quad Scenario B \quad \\ [0.5ex] 
 \hline
 Weak E.P. & $\times$ & $\times$ \\  \hline
 Strong E.P. & $\surd$ & $\times$ \\ [1ex] 
 \hline
\end{tabular}
\caption{Comparison for the validity of the Weak and Strong equivalence principles for both,  scenario A (time-diffracted free-falling particles) and scenario B (suddenly released gravitational quantum states).}
\label{table2}
\end{table}

\section*{Acknowledgements}
J.A.C.  was supported by the CONACyT master fellowship No. 725033. A.M.-R. has been partially supported by DGAPA-UNAM Project No. IA102722 and by Project CONACyT (M\'{e}xico) No. 428214.  We thank to M. Cambiaso, M. J. Everitt and C. Escobar for their careful reading of the manuscript.\\

\section*{Data Availability Statement \; } No datasets were generated or analyzed during the current study.


\bibliography{Bib.bib}

\begin{thebibliography}{45}%
\makeatletter
\providecommand \@ifxundefined [1]{%
 \@ifx{#1\undefined}
}%
\providecommand \@ifnum [1]{%
 \ifnum #1\expandafter \@firstoftwo
 \else \expandafter \@secondoftwo
 \fi
}%
\providecommand \@ifx [1]{%
 \ifx #1\expandafter \@firstoftwo
 \else \expandafter \@secondoftwo
 \fi
}%
\providecommand \natexlab [1]{#1}%
\providecommand \enquote  [1]{``#1''}%
\providecommand \bibnamefont  [1]{#1}%
\providecommand \bibfnamefont [1]{#1}%
\providecommand \citenamefont [1]{#1}%
\providecommand \href@noop [0]{\@secondoftwo}%
\providecommand \href [0]{\begingroup \@sanitize@url \@href}%
\providecommand \@href[1]{\@@startlink{#1}\@@href}%
\providecommand \@@href[1]{\endgroup#1\@@endlink}%
\providecommand \@sanitize@url [0]{\catcode `\\12\catcode `\$12\catcode
  `\&12\catcode `\#12\catcode `\^12\catcode `\_12\catcode `\%12\relax}%
\providecommand \@@startlink[1]{}%
\providecommand \@@endlink[0]{}%
\providecommand \url  [0]{\begingroup\@sanitize@url \@url }%
\providecommand \@url [1]{\endgroup\@href {#1}{\urlprefix }}%
\providecommand \urlprefix  [0]{URL }%
\providecommand \Eprint [0]{\href }%
\providecommand \doibase [0]{http://dx.doi.org/}%
\providecommand \selectlanguage [0]{\@gobble}%
\providecommand \bibinfo  [0]{\@secondoftwo}%
\providecommand \bibfield  [0]{\@secondoftwo}%
\providecommand \translation [1]{[#1]}%
\providecommand \BibitemOpen [0]{}%
\providecommand \bibitemStop [0]{}%
\providecommand \bibitemNoStop [0]{.\EOS\space}%
\providecommand \EOS [0]{\spacefactor3000\relax}%
\providecommand \BibitemShut  [1]{\csname bibitem#1\endcsname}%
\let\auto@bib@innerbib\@empty
\bibitem [{\citenamefont {Wald}(1984)}]{R_Wald_GR}%
  \BibitemOpen
  \bibfield  {author} {\bibinfo {author} {\bibfnamefont {R.~M.}\ \bibnamefont
  {Wald}},\ }\href@noop {} {\emph {\bibinfo {title} {General Relativity}}}\
  (\bibinfo  {publisher} {University of Chicago Press},\ \bibinfo {address}
  {Chicago, IL},\ \bibinfo {year} {1984})\BibitemShut {NoStop}%
\bibitem [{\citenamefont {Will}(2001)}]{C_Will}%
  \BibitemOpen
  \bibfield  {author} {\bibinfo {author} {\bibfnamefont {C.~M.}\ \bibnamefont
  {Will}},\ }\bibfield  {title} {\enquote {\bibinfo {title} {The confrontation
  between general relativity and experiment},}\ }\href {\doibase
  10.12942/lrr-2001-4} {\bibfield  {journal} {\bibinfo  {journal} {Living
  Reviews in Relativity}\ }\textbf {\bibinfo {volume} {4}},\ \bibinfo {pages}
  {4} (\bibinfo {year} {2001})}\BibitemShut {NoStop}%
\bibitem [{\citenamefont {Colella}\ \emph {et~al.}(1975)\citenamefont
  {Colella}, \citenamefont {Overhauser},\ and\ \citenamefont
  {Werner}}]{PhysRevLett.34.1472}%
  \BibitemOpen
  \bibfield  {author} {\bibinfo {author} {\bibfnamefont {R.}~\bibnamefont
  {Colella}}, \bibinfo {author} {\bibfnamefont {A.~W.}\ \bibnamefont
  {Overhauser}}, \ and\ \bibinfo {author} {\bibfnamefont {S.~A.}\ \bibnamefont
  {Werner}},\ }\bibfield  {title} {\enquote {\bibinfo {title} {Observation of
  gravitationally induced quantum interference},}\ }\href {\doibase
  10.1103/PhysRevLett.34.1472} {\bibfield  {journal} {\bibinfo  {journal}
  {Phys. Rev. Lett.}\ }\textbf {\bibinfo {volume} {34}},\ \bibinfo {pages}
  {1472--1474} (\bibinfo {year} {1975})}\BibitemShut {NoStop}%
\bibitem [{\citenamefont {Peters}\ \emph {et~al.}(1999)\citenamefont {Peters},
  \citenamefont {Chung},\ and\ \citenamefont {Chu}}]{Peters1999}%
  \BibitemOpen
  \bibfield  {author} {\bibinfo {author} {\bibfnamefont {A.}~\bibnamefont
  {Peters}}, \bibinfo {author} {\bibfnamefont {K.~Y.}\ \bibnamefont {Chung}}, \
  and\ \bibinfo {author} {\bibfnamefont {S.}~\bibnamefont {Chu}},\ }\bibfield
  {title} {\enquote {\bibinfo {title} {Measurement of gravitational
  acceleration by dropping atoms},}\ }\href {\doibase 10.1038/23655} {\bibfield
   {journal} {\bibinfo  {journal} {Nature}\ }\textbf {\bibinfo {volume}
  {400}},\ \bibinfo {pages} {849--852} (\bibinfo {year} {1999})}\BibitemShut
  {NoStop}%
\bibitem [{\citenamefont {Viola}\ and\ \citenamefont
  {Onofrio}(1997)}]{PhysRevD.55.455}%
  \BibitemOpen
  \bibfield  {author} {\bibinfo {author} {\bibfnamefont {L.}~\bibnamefont
  {Viola}}\ and\ \bibinfo {author} {\bibfnamefont {R.}~\bibnamefont
  {Onofrio}},\ }\bibfield  {title} {\enquote {\bibinfo {title} {Testing the
  equivalence principle through freely falling quantum objects},}\ }\href
  {\doibase 10.1103/PhysRevD.55.455} {\bibfield  {journal} {\bibinfo  {journal}
  {Phys. Rev. D}\ }\textbf {\bibinfo {volume} {55}},\ \bibinfo {pages}
  {455--462} (\bibinfo {year} {1997})}\BibitemShut {NoStop}%
\bibitem [{\citenamefont {Ali}\ \emph {et~al.}(2006)\citenamefont {Ali},
  \citenamefont {Majumdar}, \citenamefont {Home},\ and\ \citenamefont
  {Pan}}]{Ali_2006}%
  \BibitemOpen
  \bibfield  {author} {\bibinfo {author} {\bibfnamefont {Md~M.}\ \bibnamefont
  {Ali}}, \bibinfo {author} {\bibfnamefont {A.~S.}\ \bibnamefont {Majumdar}},
  \bibinfo {author} {\bibfnamefont {D.}~\bibnamefont {Home}}, \ and\ \bibinfo
  {author} {\bibfnamefont {A.~K.}\ \bibnamefont {Pan}},\ }\bibfield  {title}
  {\enquote {\bibinfo {title} {On the quantum analogue of
  {G}alileo{\textquotesingle}s leaning tower experiment},}\ }\href {\doibase
  10.1088/0264-9381/23/22/024} {\bibfield  {journal} {\bibinfo  {journal}
  {Classical and Quantum Gravity}\ }\textbf {\bibinfo {volume} {23}},\ \bibinfo
  {pages} {6493--6502} (\bibinfo {year} {2006})}\BibitemShut {NoStop}%
\bibitem [{\citenamefont {Chowdhury}\ \emph {et~al.}(2011)\citenamefont
  {Chowdhury}, \citenamefont {Home}, \citenamefont {Majumdar}, \citenamefont
  {Mousavi}, \citenamefont {Mozaffari},\ and\ \citenamefont
  {Sinha}}]{Chowdhury_2011}%
  \BibitemOpen
  \bibfield  {author} {\bibinfo {author} {\bibfnamefont {P.}~\bibnamefont
  {Chowdhury}}, \bibinfo {author} {\bibfnamefont {D.}~\bibnamefont {Home}},
  \bibinfo {author} {\bibfnamefont {A.~S.}\ \bibnamefont {Majumdar}}, \bibinfo
  {author} {\bibfnamefont {S.~V.}\ \bibnamefont {Mousavi}}, \bibinfo {author}
  {\bibfnamefont {M.~R.}\ \bibnamefont {Mozaffari}}, \ and\ \bibinfo {author}
  {\bibfnamefont {S.}~\bibnamefont {Sinha}},\ }\bibfield  {title} {\enquote
  {\bibinfo {title} {Strong quantum violation of the gravitational weak
  equivalence principle by a non-gaussian wave packet},}\ }\href {\doibase
  10.1088/0264-9381/29/2/025010} {\bibfield  {journal} {\bibinfo  {journal}
  {Classical and Quantum Gravity}\ }\textbf {\bibinfo {volume} {29}},\ \bibinfo
  {pages} {025010} (\bibinfo {year} {2011})}\BibitemShut {NoStop}%
\bibitem [{\citenamefont {Dumont}\ and\ \citenamefont
  {Marchioro~II}(1993)}]{PhysRevA.47.85}%
  \BibitemOpen
  \bibfield  {author} {\bibinfo {author} {\bibfnamefont {R.~S.}\ \bibnamefont
  {Dumont}}\ and\ \bibinfo {author} {\bibfnamefont {T.~L.}\ \bibnamefont
  {Marchioro~II}},\ }\bibfield  {title} {\enquote {\bibinfo {title}
  {Tunneling-time probability distribution},}\ }\href {\doibase
  10.1103/PhysRevA.47.85} {\bibfield  {journal} {\bibinfo  {journal} {Phys.
  Rev. A}\ }\textbf {\bibinfo {volume} {47}},\ \bibinfo {pages} {85--97}
  (\bibinfo {year} {1993})}\BibitemShut {NoStop}%
\bibitem [{\citenamefont {McKinnon}\ and\ \citenamefont
  {Leavens}(1995)}]{PhysRevA.51.2748}%
  \BibitemOpen
  \bibfield  {author} {\bibinfo {author} {\bibfnamefont {W.~R.}\ \bibnamefont
  {McKinnon}}\ and\ \bibinfo {author} {\bibfnamefont {C.~R.}\ \bibnamefont
  {Leavens}},\ }\bibfield  {title} {\enquote {\bibinfo {title} {Distributions
  of delay times and transmission times in {B}ohm's causal interpretation of
  quantum mechanics},}\ }\href {\doibase 10.1103/PhysRevA.51.2748} {\bibfield
  {journal} {\bibinfo  {journal} {Phys. Rev. A}\ }\textbf {\bibinfo {volume}
  {51}},\ \bibinfo {pages} {2748--2757} (\bibinfo {year} {1995})}\BibitemShut
  {NoStop}%
\bibitem [{\citenamefont {Delgado}(1999)}]{PhysRevA.59.1010}%
  \BibitemOpen
  \bibfield  {author} {\bibinfo {author} {\bibfnamefont {V.}~\bibnamefont
  {Delgado}},\ }\bibfield  {title} {\enquote {\bibinfo {title} {Quantum
  probability distribution of arrival times and probability current density},}\
  }\href {\doibase 10.1103/PhysRevA.59.1010} {\bibfield  {journal} {\bibinfo
  {journal} {Phys. Rev. A}\ }\textbf {\bibinfo {volume} {59}},\ \bibinfo
  {pages} {1010--1020} (\bibinfo {year} {1999})}\BibitemShut {NoStop}%
\bibitem [{\citenamefont {Leavens}(1998)}]{PhysRevA.58.840}%
  \BibitemOpen
  \bibfield  {author} {\bibinfo {author} {\bibfnamefont {C.~R.}\ \bibnamefont
  {Leavens}},\ }\bibfield  {title} {\enquote {\bibinfo {title} {Time of arrival
  in quantum and bohmian mechanics},}\ }\href {\doibase
  10.1103/PhysRevA.58.840} {\bibfield  {journal} {\bibinfo  {journal} {Phys.
  Rev. A}\ }\textbf {\bibinfo {volume} {58}},\ \bibinfo {pages} {840--847}
  (\bibinfo {year} {1998})}\BibitemShut {NoStop}%
\bibitem [{\citenamefont {Leavens}(1993)}]{LEAVENS199327}%
  \BibitemOpen
  \bibfield  {author} {\bibinfo {author} {\bibfnamefont {C.~R.}\ \bibnamefont
  {Leavens}},\ }\bibfield  {title} {\enquote {\bibinfo {title} {Arrival time
  distributions},}\ }\href {\doibase
  https://doi.org/10.1016/0375-9601(93)90722-C} {\bibfield  {journal} {\bibinfo
   {journal} {Physics Letters A}\ }\textbf {\bibinfo {volume} {178}},\ \bibinfo
  {pages} {27--32} (\bibinfo {year} {1993})}\BibitemShut {NoStop}%
\bibitem [{\citenamefont {Muga}\ \emph {et~al.}(1995)\citenamefont {Muga},
  \citenamefont {Brouard},\ and\ \citenamefont {Macias}}]{MUGA1995351}%
  \BibitemOpen
  \bibfield  {author} {\bibinfo {author} {\bibfnamefont {J.~G.}\ \bibnamefont
  {Muga}}, \bibinfo {author} {\bibfnamefont {S.}~\bibnamefont {Brouard}}, \
  and\ \bibinfo {author} {\bibfnamefont {D.}~\bibnamefont {Macias}},\
  }\bibfield  {title} {\enquote {\bibinfo {title} {Time of arrival in quantum
  mechanics},}\ }\href {\doibase https://doi.org/10.1006/aphy.1995.1048}
  {\bibfield  {journal} {\bibinfo  {journal} {Annals of Physics}\ }\textbf
  {\bibinfo {volume} {240}},\ \bibinfo {pages} {351--366} (\bibinfo {year}
  {1995})}\BibitemShut {NoStop}%
\bibitem [{\citenamefont {Davies}(2004{\natexlab{a}})}]{Davies_2004}%
  \BibitemOpen
  \bibfield  {author} {\bibinfo {author} {\bibfnamefont {P.~C.~W.}\
  \bibnamefont {Davies}},\ }\bibfield  {title} {\enquote {\bibinfo {title}
  {Quantum mechanics and the equivalence principle},}\ }\href {\doibase
  10.1088/0264-9381/21/11/017} {\bibfield  {journal} {\bibinfo  {journal}
  {Classical and Quantum Gravity}\ }\textbf {\bibinfo {volume} {21}},\ \bibinfo
  {pages} {2761--2772} (\bibinfo {year} {2004}{\natexlab{a}})}\BibitemShut
  {NoStop}%
\bibitem [{\citenamefont {Davies}(2004{\natexlab{b}})}]{Davies2_2004}%
  \BibitemOpen
  \bibfield  {author} {\bibinfo {author} {\bibfnamefont {P.~C.~W.}\
  \bibnamefont {Davies}},\ }\bibfield  {title} {\enquote {\bibinfo {title}
  {Transit time of a freely falling quantum particle in a background
  gravitational field},}\ }\href {\doibase 10.1088/0264-9381/21/24/001}
  {\bibfield  {journal} {\bibinfo  {journal} {Classical and Quantum Gravity}\
  }\textbf {\bibinfo {volume} {21}},\ \bibinfo {pages} {5677--5683} (\bibinfo
  {year} {2004}{\natexlab{b}})}\BibitemShut {NoStop}%
\bibitem [{\citenamefont {Salecker}\ and\ \citenamefont
  {Wigner}(1958)}]{PhysRev.109.571}%
  \BibitemOpen
  \bibfield  {author} {\bibinfo {author} {\bibfnamefont {H.}~\bibnamefont
  {Salecker}}\ and\ \bibinfo {author} {\bibfnamefont {E.~P.}\ \bibnamefont
  {Wigner}},\ }\bibfield  {title} {\enquote {\bibinfo {title} {Quantum
  limitations of the measurement of space-time distances},}\ }\href {\doibase
  10.1103/PhysRev.109.571} {\bibfield  {journal} {\bibinfo  {journal} {Phys.
  Rev.}\ }\textbf {\bibinfo {volume} {109}},\ \bibinfo {pages} {571--577}
  (\bibinfo {year} {1958})}\BibitemShut {NoStop}%
\bibitem [{\citenamefont {Peres}(1980)}]{doi:10.1119/1.12061}%
  \BibitemOpen
  \bibfield  {author} {\bibinfo {author} {\bibfnamefont {A.}~\bibnamefont
  {Peres}},\ }\bibfield  {title} {\enquote {\bibinfo {title} {Measurement of
  time by quantum clocks},}\ }\href {\doibase 10.1119/1.12061} {\bibfield
  {journal} {\bibinfo  {journal} {American Journal of Physics}\ }\textbf
  {\bibinfo {volume} {48}},\ \bibinfo {pages} {552--557} (\bibinfo {year}
  {1980})}\BibitemShut {NoStop}%
\bibitem [{\citenamefont {Anastopoulos}\ and\ \citenamefont
  {Hu}(2018)}]{Anastopoulos_2018}%
  \BibitemOpen
  \bibfield  {author} {\bibinfo {author} {\bibfnamefont {C.}~\bibnamefont
  {Anastopoulos}}\ and\ \bibinfo {author} {\bibfnamefont {B.~L.}\ \bibnamefont
  {Hu}},\ }\bibfield  {title} {\enquote {\bibinfo {title} {Equivalence
  principle for quantum systems: dephasing and phase shift of free-falling
  particles},}\ }\href {\doibase 10.1088/1361-6382/aaa0e8} {\bibfield
  {journal} {\bibinfo  {journal} {Classical and Quantum Gravity}\ }\textbf
  {\bibinfo {volume} {35}},\ \bibinfo {pages} {035011} (\bibinfo {year}
  {2018})}\BibitemShut {NoStop}%
\bibitem [{\citenamefont {Finkelstein}(1999)}]{PhysRevA.59.3218}%
  \BibitemOpen
  \bibfield  {author} {\bibinfo {author} {\bibfnamefont {J.}~\bibnamefont
  {Finkelstein}},\ }\bibfield  {title} {\enquote {\bibinfo {title} {Ambiguities
  of arrival-time distributions in quantum theory},}\ }\href {\doibase
  10.1103/PhysRevA.59.3218} {\bibfield  {journal} {\bibinfo  {journal} {Phys.
  Rev. A}\ }\textbf {\bibinfo {volume} {59}},\ \bibinfo {pages} {3218--3222}
  (\bibinfo {year} {1999})}\BibitemShut {NoStop}%
\bibitem [{\citenamefont {Muga}\ and\ \citenamefont
  {Leavens}(2000)}]{MUGA2000353}%
  \BibitemOpen
  \bibfield  {author} {\bibinfo {author} {\bibfnamefont {J.~G.}\ \bibnamefont
  {Muga}}\ and\ \bibinfo {author} {\bibfnamefont {C.~R.}\ \bibnamefont
  {Leavens}},\ }\bibfield  {title} {\enquote {\bibinfo {title} {Arrival time in
  quantum mechanics},}\ }\href {\doibase
  https://doi.org/10.1016/S0370-1573(00)00047-8} {\bibfield  {journal}
  {\bibinfo  {journal} {Physics Reports}\ }\textbf {\bibinfo {volume} {338}},\
  \bibinfo {pages} {353--438} (\bibinfo {year} {2000})}\BibitemShut {NoStop}%
\bibitem [{\citenamefont {Moshinsky}(1952)}]{PhysRev.88.625}%
  \BibitemOpen
  \bibfield  {author} {\bibinfo {author} {\bibfnamefont {M.}~\bibnamefont
  {Moshinsky}},\ }\bibfield  {title} {\enquote {\bibinfo {title} {Diffraction
  in time},}\ }\href {\doibase 10.1103/PhysRev.88.625} {\bibfield  {journal}
  {\bibinfo  {journal} {Phys. Rev.}\ }\textbf {\bibinfo {volume} {88}},\
  \bibinfo {pages} {625--631} (\bibinfo {year} {1952})}\BibitemShut {NoStop}%
\bibitem [{\citenamefont {Szriftgiser}\ \emph {et~al.}(1996)\citenamefont
  {Szriftgiser}, \citenamefont {Gu\'ery-Odelin}, \citenamefont {Arndt},\ and\
  \citenamefont {Dalibard}}]{PhysRevLett.77.4}%
  \BibitemOpen
  \bibfield  {author} {\bibinfo {author} {\bibfnamefont {P.}~\bibnamefont
  {Szriftgiser}}, \bibinfo {author} {\bibfnamefont {D.}~\bibnamefont
  {Gu\'ery-Odelin}}, \bibinfo {author} {\bibfnamefont {M.}~\bibnamefont
  {Arndt}}, \ and\ \bibinfo {author} {\bibfnamefont {J.}~\bibnamefont
  {Dalibard}},\ }\bibfield  {title} {\enquote {\bibinfo {title} {Atomic wave
  diffraction and interference using temporal slits},}\ }\href {\doibase
  10.1103/PhysRevLett.77.4} {\bibfield  {journal} {\bibinfo  {journal} {Phys.
  Rev. Lett.}\ }\textbf {\bibinfo {volume} {77}},\ \bibinfo {pages} {4--7}
  (\bibinfo {year} {1996})}\BibitemShut {NoStop}%
\bibitem [{\citenamefont {{del Campo}}\ \emph {et~al.}(2009)\citenamefont {{del
  Campo}}, \citenamefont {García-Calderón},\ and\ \citenamefont
  {Muga}}]{DELCAMPO20091}%
  \BibitemOpen
  \bibfield  {author} {\bibinfo {author} {\bibfnamefont {A.}~\bibnamefont {{del
  Campo}}}, \bibinfo {author} {\bibfnamefont {G.}~\bibnamefont
  {García-Calderón}}, \ and\ \bibinfo {author} {\bibfnamefont {J.~G.}\
  \bibnamefont {Muga}},\ }\bibfield  {title} {\enquote {\bibinfo {title}
  {Quantum transients},}\ }\href {\doibase
  https://doi.org/10.1016/j.physrep.2009.03.002} {\bibfield  {journal}
  {\bibinfo  {journal} {Physics Reports}\ }\textbf {\bibinfo {volume} {476}},\
  \bibinfo {pages} {1--50} (\bibinfo {year} {2009})}\BibitemShut {NoStop}%
\bibitem [{\citenamefont {Das}\ and\ \citenamefont
  {Vagenas}(2008)}]{PhysRevLett.101.221301}%
  \BibitemOpen
  \bibfield  {author} {\bibinfo {author} {\bibfnamefont {Saurya}\ \bibnamefont
  {Das}}\ and\ \bibinfo {author} {\bibfnamefont {Elias~C.}\ \bibnamefont
  {Vagenas}},\ }\bibfield  {title} {\enquote {\bibinfo {title} {Universality of
  quantum gravity corrections},}\ }\href {\doibase
  10.1103/PhysRevLett.101.221301} {\bibfield  {journal} {\bibinfo  {journal}
  {Phys. Rev. Lett.}\ }\textbf {\bibinfo {volume} {101}},\ \bibinfo {pages}
  {221301} (\bibinfo {year} {2008})}\BibitemShut {NoStop}%
\bibitem [{\citenamefont {Mart\'{\i}n-Ruiz}(2014)}]{PhysRevD.90.125027}%
  \BibitemOpen
  \bibfield  {author} {\bibinfo {author} {\bibfnamefont {A.}~\bibnamefont
  {Mart\'{\i}n-Ruiz}},\ }\bibfield  {title} {\enquote {\bibinfo {title}
  {Diffraction in time of polymer particles},}\ }\href {\doibase
  10.1103/PhysRevD.90.125027} {\bibfield  {journal} {\bibinfo  {journal} {Phys.
  Rev. D}\ }\textbf {\bibinfo {volume} {90}},\ \bibinfo {pages} {125027}
  (\bibinfo {year} {2014})}\BibitemShut {NoStop}%
\bibitem [{\citenamefont {Longhi}(2018)}]{Longhi:18}%
  \BibitemOpen
  \bibfield  {author} {\bibinfo {author} {\bibfnamefont {S.}~\bibnamefont
  {Longhi}},\ }\bibfield  {title} {\enquote {\bibinfo {title} {Equivalence
  principle and quantum mechanics: quantum simulation with entangled
  photons},}\ }\href {\doibase 10.1364/OL.43.000226} {\bibfield  {journal}
  {\bibinfo  {journal} {Opt. Lett.}\ }\textbf {\bibinfo {volume} {43}},\
  \bibinfo {pages} {226--229} (\bibinfo {year} {2018})}\BibitemShut {NoStop}%
\bibitem [{\citenamefont {Feynman}\ and\ \citenamefont
  {A.}(1965)}]{Feynman_Hibbs}%
  \BibitemOpen
  \bibfield  {author} {\bibinfo {author} {\bibfnamefont {R.P.}\ \bibnamefont
  {Feynman}}\ and\ \bibinfo {author} {\bibfnamefont {Hibbs}\ \bibnamefont
  {A.}},\ }\href {https://cds.cern.ch/record/100771} {\emph {\bibinfo {title}
  {{Quantum mechanics and path integrals}}}},\ International series in pure and
  applied physics\ (\bibinfo  {publisher} {McGraw-Hill},\ \bibinfo {address}
  {New York, NY},\ \bibinfo {year} {1965})\BibitemShut {NoStop}%
\bibitem [{\citenamefont {Gradshteyn}\ and\ \citenamefont
  {Ryzhik}(1994)}]{Gradshteyn_Ryzhik}%
  \BibitemOpen
  \bibfield  {author} {\bibinfo {author} {\bibfnamefont {I.~S.}\ \bibnamefont
  {Gradshteyn}}\ and\ \bibinfo {author} {\bibfnamefont {I.~M.}\ \bibnamefont
  {Ryzhik}},\ }\href {10.7208/chicago/9780226870373.001.0001} {\emph {\bibinfo
  {title} {Table of Integrals, Series, and Products}}},\ \bibinfo {edition}
  {4th}\ ed.\ (\bibinfo  {publisher} {Academic Press},\ \bibinfo {address} {New
  York},\ \bibinfo {year} {1994})\ \bibinfo {note} {edited by A. Jeffrey and D.
  Zwillinger}\BibitemShut {NoStop}%
\bibitem [{\citenamefont {Emrich}(2016)}]{EMRICH201655}%
  \BibitemOpen
  \bibfield  {author} {\bibinfo {author} {\bibfnamefont {William}\ \bibnamefont
  {Emrich}},\ }\bibfield  {title} {\enquote {\bibinfo {title} {Chapter 5 -
  basic nuclear structure and processes},}\ }in\ \href {\doibase
  https://doi.org/10.1016/B978-0-12-804474-2.00005-9} {\emph {\bibinfo
  {booktitle} {Principles of Nuclear Rocket Propulsion}}},\ \bibinfo {editor}
  {edited by\ \bibinfo {editor} {\bibfnamefont {William}\ \bibnamefont
  {Emrich}}}\ (\bibinfo  {publisher} {Butterworth-Heinemann},\ \bibinfo {year}
  {2016})\ pp.\ \bibinfo {pages} {55--80}\BibitemShut {NoStop}%
\bibitem [{\citenamefont {Nesvizhevsky}\ \emph {et~al.}(2002)\citenamefont
  {Nesvizhevsky}, \citenamefont {Börner}, \citenamefont {Petukhov},
  \citenamefont {Abele}, \citenamefont {Baeßler}, \citenamefont {Rueß},
  \citenamefont {Stöferle}, \citenamefont {Westphal}, \citenamefont
  {Gagarski}, \citenamefont {Petrov},\ and\ \citenamefont
  {Strelkov}}]{Nesvizhevsky2002}%
  \BibitemOpen
  \bibfield  {author} {\bibinfo {author} {\bibfnamefont {V.~V.}\ \bibnamefont
  {Nesvizhevsky}}, \bibinfo {author} {\bibfnamefont {H.~G.}\ \bibnamefont
  {Börner}}, \bibinfo {author} {\bibfnamefont {A.~K.}\ \bibnamefont
  {Petukhov}}, \bibinfo {author} {\bibfnamefont {H.}~\bibnamefont {Abele}},
  \bibinfo {author} {\bibfnamefont {S.}~\bibnamefont {Baeßler}}, \bibinfo
  {author} {\bibfnamefont {F.~J.}\ \bibnamefont {Rueß}}, \bibinfo {author}
  {\bibfnamefont {T.}~\bibnamefont {Stöferle}}, \bibinfo {author}
  {\bibfnamefont {A.}~\bibnamefont {Westphal}}, \bibinfo {author}
  {\bibfnamefont {A.~M.}\ \bibnamefont {Gagarski}}, \bibinfo {author}
  {\bibfnamefont {G.~A.}\ \bibnamefont {Petrov}}, \ and\ \bibinfo {author}
  {\bibfnamefont {A.~V.}\ \bibnamefont {Strelkov}},\ }\bibfield  {title}
  {\enquote {\bibinfo {title} {Quantum states of neutrons in the earth's
  gravitational field},}\ }\href {\doibase 10.1038/415297a} {\bibfield
  {journal} {\bibinfo  {journal} {Nature}\ }\textbf {\bibinfo {volume} {415}},\
  \bibinfo {pages} {297--299} (\bibinfo {year} {2002})}\BibitemShut {NoStop}%
\bibitem [{\citenamefont {Aminoff}\ \emph {et~al.}(1993)\citenamefont
  {Aminoff}, \citenamefont {Steane}, \citenamefont {Bouyer}, \citenamefont
  {Desbiolles}, \citenamefont {Dalibard},\ and\ \citenamefont
  {Cohen-Tannoudji}}]{PhysRevLett.71.3083}%
  \BibitemOpen
  \bibfield  {author} {\bibinfo {author} {\bibfnamefont {C.~G.}\ \bibnamefont
  {Aminoff}}, \bibinfo {author} {\bibfnamefont {A.~M.}\ \bibnamefont {Steane}},
  \bibinfo {author} {\bibfnamefont {P.}~\bibnamefont {Bouyer}}, \bibinfo
  {author} {\bibfnamefont {P.}~\bibnamefont {Desbiolles}}, \bibinfo {author}
  {\bibfnamefont {J.}~\bibnamefont {Dalibard}}, \ and\ \bibinfo {author}
  {\bibfnamefont {C.}~\bibnamefont {Cohen-Tannoudji}},\ }\bibfield  {title}
  {\enquote {\bibinfo {title} {Cesium atoms bouncing in a stable gravitational
  cavity},}\ }\href {\doibase 10.1103/PhysRevLett.71.3083} {\bibfield
  {journal} {\bibinfo  {journal} {Phys. Rev. Lett.}\ }\textbf {\bibinfo
  {volume} {71}},\ \bibinfo {pages} {3083--3086} (\bibinfo {year}
  {1993})}\BibitemShut {NoStop}%
\bibitem [{\citenamefont {Arndt}\ \emph {et~al.}(1999)\citenamefont {Arndt},
  \citenamefont {Nairz}, \citenamefont {Vos-Andreae}, \citenamefont {Keller},
  \citenamefont {van~der Zouw},\ and\ \citenamefont {Zeilinger}}]{Arndt1999}%
  \BibitemOpen
  \bibfield  {author} {\bibinfo {author} {\bibfnamefont {M.}~\bibnamefont
  {Arndt}}, \bibinfo {author} {\bibfnamefont {O.}~\bibnamefont {Nairz}},
  \bibinfo {author} {\bibfnamefont {J.}~\bibnamefont {Vos-Andreae}}, \bibinfo
  {author} {\bibfnamefont {C.}~\bibnamefont {Keller}}, \bibinfo {author}
  {\bibfnamefont {G.}~\bibnamefont {van~der Zouw}}, \ and\ \bibinfo {author}
  {\bibfnamefont {A.}~\bibnamefont {Zeilinger}},\ }\bibfield  {title} {\enquote
  {\bibinfo {title} {Wave–particle duality of {C}60 molecules},}\ }\href
  {\doibase 10.1038/44348} {\bibfield  {journal} {\bibinfo  {journal} {Nature}\
  }\textbf {\bibinfo {volume} {401}},\ \bibinfo {pages} {680--682} (\bibinfo
  {year} {1999})}\BibitemShut {NoStop}%
\bibitem [{\citenamefont {Goel}\ \emph {et~al.}(2004)\citenamefont {Goel},
  \citenamefont {Howard},\ and\ \citenamefont {Vander~Sande}}]{Goel2004}%
  \BibitemOpen
  \bibfield  {author} {\bibinfo {author} {\bibfnamefont {A.}~\bibnamefont
  {Goel}}, \bibinfo {author} {\bibfnamefont {J.~B.}\ \bibnamefont {Howard}}, \
  and\ \bibinfo {author} {\bibfnamefont {J.~B}\ \bibnamefont {Vander~Sande}},\
  }\bibfield  {title} {\enquote {\bibinfo {title} {Size analysis of single
  fullerene molecules by electron microscopy},}\ }\href {\doibase
  10.1016/j.carbon.2004.03.022} {\bibfield  {journal} {\bibinfo  {journal}
  {Carbon}\ }\textbf {\bibinfo {volume} {42}},\ \bibinfo {pages} {1907--1915}
  (\bibinfo {year} {2004})}\BibitemShut {NoStop}%
\bibitem [{\citenamefont {Robinett}(1995)}]{doi:10.1119/1.17807}%
  \BibitemOpen
  \bibfield  {author} {\bibinfo {author} {\bibfnamefont {R.~W.}\ \bibnamefont
  {Robinett}},\ }\bibfield  {title} {\enquote {\bibinfo {title} {Quantum and
  classical probability distributions for position and momentum},}\ }\href
  {\doibase 10.1119/1.17807} {\bibfield  {journal} {\bibinfo  {journal}
  {American Journal of Physics}\ }\textbf {\bibinfo {volume} {63}},\ \bibinfo
  {pages} {823--832} (\bibinfo {year} {1995})}\BibitemShut {NoStop}%
\bibitem [{\citenamefont {Yoder}(2006)}]{doi:10.1119/1.2173280}%
  \BibitemOpen
  \bibfield  {author} {\bibinfo {author} {\bibfnamefont {G.}~\bibnamefont
  {Yoder}},\ }\bibfield  {title} {\enquote {\bibinfo {title} {Using classical
  probability functions to illuminate the relation between classical and
  quantum physics},}\ }\href {\doibase 10.1119/1.2173280} {\bibfield  {journal}
  {\bibinfo  {journal} {American Journal of Physics}\ }\textbf {\bibinfo
  {volume} {74}},\ \bibinfo {pages} {404--411} (\bibinfo {year}
  {2006})}\BibitemShut {NoStop}%
\bibitem [{\citenamefont {Rowe}(1987)}]{Rowe_1987}%
  \BibitemOpen
  \bibfield  {author} {\bibinfo {author} {\bibfnamefont {E~G~P}\ \bibnamefont
  {Rowe}},\ }\bibfield  {title} {\enquote {\bibinfo {title} {The classical
  limit of quantum mechanical hydrogen radial distributions},}\ }\href
  {\doibase 10.1088/0143-0807/8/2/002} {\bibfield  {journal} {\bibinfo
  {journal} {European Journal of Physics}\ }\textbf {\bibinfo {volume} {8}},\
  \bibinfo {pages} {81--87} (\bibinfo {year} {1987})}\BibitemShut {NoStop}%
\bibitem [{\citenamefont {Bernal}\ \emph {et~al.}(2013)\citenamefont {Bernal},
  \citenamefont {Mart\'{\i}n-Ruiz},\ and\ \citenamefont
  {García-Melgarejo}}]{ClassLim1}%
  \BibitemOpen
  \bibfield  {author} {\bibinfo {author} {\bibfnamefont {J.}~\bibnamefont
  {Bernal}}, \bibinfo {author} {\bibfnamefont {A.}~\bibnamefont
  {Mart\'{\i}n-Ruiz}}, \ and\ \bibinfo {author} {\bibfnamefont
  {J.}~\bibnamefont {García-Melgarejo}},\ }\bibfield  {title} {\enquote
  {\bibinfo {title} {A simple mathematical formulation of the correspondence
  principle},}\ }\href {\doibase 10.4236/jmp.2013.41017} {\bibfield  {journal}
  {\bibinfo  {journal} {Journal of Modern Physics,}\ }\textbf {\bibinfo
  {volume} {4}},\ \bibinfo {pages} {108} (\bibinfo {year} {2013})}\BibitemShut
  {NoStop}%
\bibitem [{\citenamefont {Mart\'{\i}n-Ruiz}\ \emph
  {et~al.}(2013{\natexlab{a}})\citenamefont {Mart\'{\i}n-Ruiz}, \citenamefont
  {Bernal}, \citenamefont {Frank},\ and\ \citenamefont
  {Carbajal-Dominguez}}]{ClassLim2}%
  \BibitemOpen
  \bibfield  {author} {\bibinfo {author} {\bibfnamefont {A.}~\bibnamefont
  {Mart\'{\i}n-Ruiz}}, \bibinfo {author} {\bibfnamefont {J.}~\bibnamefont
  {Bernal}}, \bibinfo {author} {\bibfnamefont {A.}~\bibnamefont {Frank}}, \
  and\ \bibinfo {author} {\bibfnamefont {A.}~\bibnamefont
  {Carbajal-Dominguez}},\ }\bibfield  {title} {\enquote {\bibinfo {title} {The
  classical limit of the quantum {K}epler problem},}\ }\href {\doibase
  10.4236/jmp.2013.46112} {\bibfield  {journal} {\bibinfo  {journal} {Journal
  of Modern Physics,}\ }\textbf {\bibinfo {volume} {4}},\ \bibinfo {pages}
  {818} (\bibinfo {year} {2013}{\natexlab{a}})}\BibitemShut {NoStop}%
\bibitem [{\citenamefont {Mart\'{\i}n-Ruiz}\ \emph
  {et~al.}(2013{\natexlab{b}})\citenamefont {Mart\'{\i}n-Ruiz}, \citenamefont
  {Bernal},\ and\ \citenamefont {Carbajal-Dominguez}}]{ClassLim3}%
  \BibitemOpen
  \bibfield  {author} {\bibinfo {author} {\bibfnamefont {A.}~\bibnamefont
  {Mart\'{\i}n-Ruiz}}, \bibinfo {author} {\bibfnamefont {J.}~\bibnamefont
  {Bernal}}, \ and\ \bibinfo {author} {\bibfnamefont {A.}~\bibnamefont
  {Carbajal-Dominguez}},\ }\bibfield  {title} {\enquote {\bibinfo {title}
  {Macroscopic quantum behaviour of periodic quantum systems},}\ }\href
  {\doibase 10.4236/jmp.2014.51007} {\bibfield  {journal} {\bibinfo  {journal}
  {Journal of Modern Physics,}\ }\textbf {\bibinfo {volume} {5}},\ \bibinfo
  {pages} {44} (\bibinfo {year} {2013}{\natexlab{b}})}\BibitemShut {NoStop}%
\bibitem [{\citenamefont {Ca\~{n}as}\ \emph {et~al.}()\citenamefont
  {Ca\~{n}as}, \citenamefont {Bernal},\ and\ \citenamefont
  {Mart\'{i}n-Ruiz}}]{Nuevo}%
  \BibitemOpen
  \bibfield  {author} {\bibinfo {author} {\bibfnamefont {Juan~A.}\ \bibnamefont
  {Ca\~{n}as}}, \bibinfo {author} {\bibfnamefont {J.}~\bibnamefont {Bernal}}, \
  and\ \bibinfo {author} {\bibfnamefont {A.}~\bibnamefont {Mart\'{i}n-Ruiz}},\
  }\href@noop {} {\enquote {\bibinfo {title} {Exact classical limit of the
  quantum bouncer},}\ }\bibinfo {note} {Unpublished}\BibitemShut {NoStop}%
\bibitem [{\citenamefont {Baeßler}\ \emph {et~al.}(2009)\citenamefont
  {Baeßler}, \citenamefont {Nesvizhevsky}, \citenamefont {Pignol},
  \citenamefont {Protasov},\ and\ \citenamefont {Voronin}}]{BAELER2009149}%
  \BibitemOpen
  \bibfield  {author} {\bibinfo {author} {\bibfnamefont {S.}~\bibnamefont
  {Baeßler}}, \bibinfo {author} {\bibfnamefont {V.~V.}\ \bibnamefont
  {Nesvizhevsky}}, \bibinfo {author} {\bibfnamefont {G.}~\bibnamefont
  {Pignol}}, \bibinfo {author} {\bibfnamefont {K.~V.}\ \bibnamefont
  {Protasov}}, \ and\ \bibinfo {author} {\bibfnamefont {A.~Yu.}\ \bibnamefont
  {Voronin}},\ }\bibfield  {title} {\enquote {\bibinfo {title} {Constraints on
  spin-dependent short-range interactions using gravitational quantum levels of
  ultracold neutrons},}\ }\href {\doibase
  https://doi.org/10.1016/j.nima.2009.07.048} {\bibfield  {journal} {\bibinfo
  {journal} {Nuclear Instruments and Methods in Physics Research Section A:
  Accelerators, Spectrometers, Detectors and Associated Equipment}\ }\textbf
  {\bibinfo {volume} {611}},\ \bibinfo {pages} {149--152} (\bibinfo {year}
  {2009})},\ \bibinfo {note} {particle Physics with Slow Neutrons}\BibitemShut
  {NoStop}%
\bibitem [{\citenamefont {Mart\'{\i}n-Ruiz}\ and\ \citenamefont
  {Escobar}(2018)}]{PhysRevD.97.095039}%
  \BibitemOpen
  \bibfield  {author} {\bibinfo {author} {\bibfnamefont {A.}~\bibnamefont
  {Mart\'{\i}n-Ruiz}}\ and\ \bibinfo {author} {\bibfnamefont {C.~A.}\
  \bibnamefont {Escobar}},\ }\bibfield  {title} {\enquote {\bibinfo {title}
  {Testing {L}orentz and {CPT} invariance with ultracold neutrons},}\ }\href
  {\doibase 10.1103/PhysRevD.97.095039} {\bibfield  {journal} {\bibinfo
  {journal} {Phys. Rev. D}\ }\textbf {\bibinfo {volume} {97}},\ \bibinfo
  {pages} {095039} (\bibinfo {year} {2018})}\BibitemShut {NoStop}%
\bibitem [{\citenamefont {Escobar}\ and\ \citenamefont
  {Mart\'{\i}n-Ruiz}(2019)}]{PhysRevD.99.075032}%
  \BibitemOpen
  \bibfield  {author} {\bibinfo {author} {\bibfnamefont {C.~A.}\ \bibnamefont
  {Escobar}}\ and\ \bibinfo {author} {\bibfnamefont {A.}~\bibnamefont
  {Mart\'{\i}n-Ruiz}},\ }\bibfield  {title} {\enquote {\bibinfo {title}
  {Gravitational searches for {L}orentz violation with ultracold neutrons},}\
  }\href {\doibase 10.1103/PhysRevD.99.075032} {\bibfield  {journal} {\bibinfo
  {journal} {Phys. Rev. D}\ }\textbf {\bibinfo {volume} {99}},\ \bibinfo
  {pages} {075032} (\bibinfo {year} {2019})}\BibitemShut {NoStop}%
\bibitem [{\citenamefont {Vallée}\ and\ \citenamefont
  {Soares}(2004)}]{doi:10.1142/p345}%
  \BibitemOpen
  \bibfield  {author} {\bibinfo {author} {\bibfnamefont {O.}~\bibnamefont
  {Vallée}}\ and\ \bibinfo {author} {\bibfnamefont {M.}~\bibnamefont
  {Soares}},\ }\href {\doibase 10.1142/p345} {\emph {\bibinfo {title} {Airy
  Functions and Applications to Physics}}}\ (\bibinfo  {publisher} {IMPERIAL
  COLLEGE PRESS},\ \bibinfo {year} {2004})\ \Eprint
  {http://arxiv.org/abs/https://www.worldscientific.com/doi/pdf/10.1142/p345}
  {https://www.worldscientific.com/doi/pdf/10.1142/p345} \BibitemShut {NoStop}%
\bibitem [{\citenamefont {Nesvizhevsky}\ \emph {et~al.}(2000)\citenamefont
  {Nesvizhevsky}, \citenamefont {Börner}, \citenamefont {Gagarski},
  \citenamefont {Petrov}, \citenamefont {Petukhov}, \citenamefont {Abele},
  \citenamefont {Bäßler}, \citenamefont {Stöferle},\ and\ \citenamefont
  {Soloviev}}]{NESVIZHEVSKY2000754}%
  \BibitemOpen
  \bibfield  {author} {\bibinfo {author} {\bibfnamefont {V.~V.}\ \bibnamefont
  {Nesvizhevsky}}, \bibinfo {author} {\bibfnamefont {H.}~\bibnamefont
  {Börner}}, \bibinfo {author} {\bibfnamefont {A.~M.}\ \bibnamefont
  {Gagarski}}, \bibinfo {author} {\bibfnamefont {G.~A.}\ \bibnamefont
  {Petrov}}, \bibinfo {author} {\bibfnamefont {A.~K.}\ \bibnamefont
  {Petukhov}}, \bibinfo {author} {\bibfnamefont {H.}~\bibnamefont {Abele}},
  \bibinfo {author} {\bibfnamefont {S.}~\bibnamefont {Bäßler}}, \bibinfo
  {author} {\bibfnamefont {T.}~\bibnamefont {Stöferle}}, \ and\ \bibinfo
  {author} {\bibfnamefont {S.~M.}\ \bibnamefont {Soloviev}},\ }\bibfield
  {title} {\enquote {\bibinfo {title} {Search for quantum states of the neutron
  in a gravitational field: gravitational levels},}\ }\href {\doibase
  https://doi.org/10.1016/S0168-9002(99)01077-3} {\bibfield  {journal}
  {\bibinfo  {journal} {Nuclear Instruments and Methods in Physics Research
  Section A: Accelerators, Spectrometers, Detectors and Associated Equipment}\
  }\textbf {\bibinfo {volume} {440}},\ \bibinfo {pages} {754--759} (\bibinfo
  {year} {2000})}\BibitemShut {NoStop}%
\end{thebibliography}%

\end{document}